**Geophysics-informed neural network for model-based seismic inversion using surrogate point spread functions**

M. Saraiva[1], A. Muller[1], A. Maul[1]
[1] Petrobras

**Introduction**

Model-based seismic inversion is a cornerstone of seismic reservoir characterization, widely used to estimate porosity trends (Simm & Bacon, 2014). This iterative process updates a seismic impedance model by minimizing the mismatch between observed and modelled seismic data (Russel, 1988). However, traditional approaches rely on 1D stationary average wavelets for seismic modeling, which lack lateral resolution as they fail to account for realistic seismic lateral variations caused by acquisition geometries and overburden complexities. This simplification results in suboptimal acoustic impedance approximations, compromising the accuracy and reliability of inversion results. These limitations highlight the need for innovative methodologies to address the inherent shortcomings of 1D wavelets.

An alternative to 1D stationary wavelets is the use of Point Spread Functions (PSFs) in seismic modeling (Lecomte, 2008). PSFs can be viewed as the impulse response of seismic acquisition and processing. They are spatially varying representations of seismic resolution that account for finite seismic bandwidth and non-stationary behavior. These characteristics address the primary limitations of 1D stationary wavelets by providing a more realistic depiction of seismic data, particularly in regions with complex geological features. Despite these advantages, the adoption of PSFs presents challenges: estimating a full PSFs volume requires specialized seismic processing workflows, which are both computationally expensive and time-intensive in terms of computation and human effort.

To overcome these challenges, we propose a novel approach that leverages a Deep Convolutional Neural Network (DCNN) to simultaneously generate a surrogate for the PSF and predict the acoustic impedance (IP). The network's output is used to calculate reflection coefficients (RC), which are subsequently convolved with the generated PSF to simulate seismic amplitude data. This embedded 2D convolutional seismic modeling enables the formulation of a loss function that compares the modeled seismic amplitude with the input (true) seismic amplitude. By backpropagating this error through the network, the DCNN iteratively updates its weights to generate RC and PSF estimates that better fit the observed seismic data. This methodology draws inspiration from Müller et al. (2023), which employs Physics-Informed Neural Networks (PINNs) to estimate seismic interval velocity. However, instead of embedding explicit physical laws directly into the loss function, we incorporate convolutional seismic modeling processes rooted in geophysical principles. Consequently, we term this architecture a Geophysics-Informed Neural Network (GINN), reflecting its tailored integration of seismic inversion principles and deep learning techniques.

**Methodology**

For this experiment, we used synthetic data, a crop of a central inline extracted from the SEAM Phase I Earth Model, as shown in Figure 1(a). This model is widely recognized for its realistic representation of subsurface geology, including complex structures such as salt bodies, faults, and stratigraphic variations, making it a suitable benchmark for seismic inversion experiments. Based on this inline, seismic shots were modeled on a 2D grid and subsequently migrated using Reverse Time Migration (RTM), ensuring high-resolution imaging of subsurface structures, as illustrated in Figure 1(b).

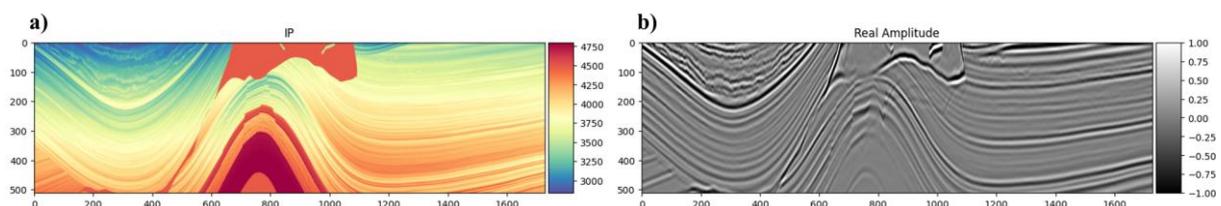

*Figure 1* A crop of a central inline extracted from the SEAM Phase I Earth Model: a) Synthetic IP model, b) Reverse Time Migration (RTM) seismic amplitude.

The resulting RTM seismic amplitude image serves as the first input for our GINN. To enhance the network's ability to account for depth and lateral positioning, additional positional information for each pixel within the seismic grid was included as auxiliary input features. This integration allows the GINN to better handle variations in resolution and amplitude related to changes in depth and lateral position (Saraiva et al., 2021).

During the training process, the low-frequency impedance (LF-IP) model was incorporated into the seismic modeling stage. Including LF- IP information is a common practice in inversion processes, as seismic data lacks low-frequency content critical for reconstructing absolute impedance contrasts (Claerbout, 1985).

The dataset for training the GINN was created by extracting sequential patches of size $128 \times 128$ samples, with an overlap of $64 \times 64$ samples, using the TorchIO library. The patches were shuffled to improve training dynamics.

We constructed the network using the PyTorch library, implementing a 2D UNet architecture. The UNet design was chosen for its proven effectiveness in tasks involving image-to-image transformations (Ronneberger et al., 2015). The network's encoder was built using ResNet modules while the decoder comprises upsampling and Conv2D blocks, designed to reconstruct outputs at the original resolution while preserving spatial details. The GINN produces three outputs: pseudo-impedance (pseudo-IP), proto zero-phase PSF, and residual PSF (Figure 2).

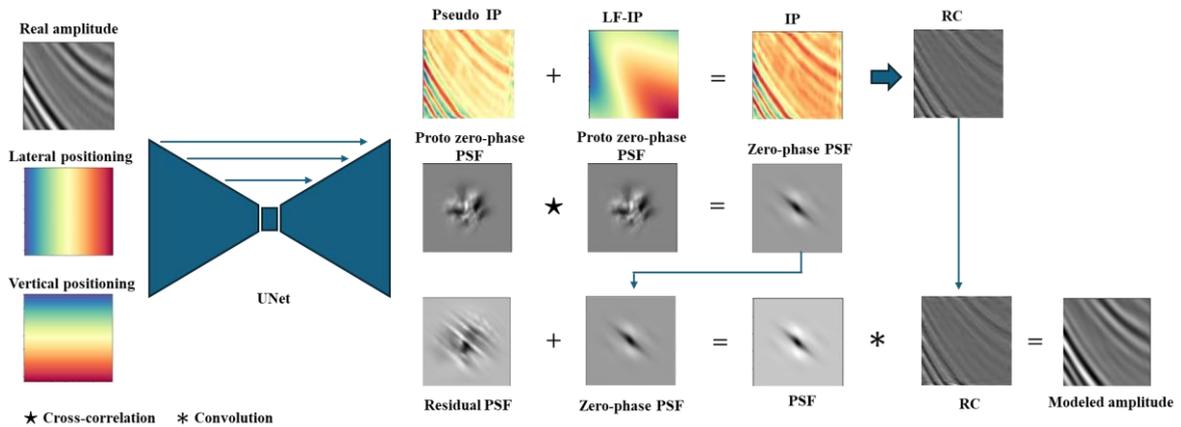

*Figure 2* GINN pipeline illustrating the inputs (real amplitude, lateral and vertical positioning patches), the UNet architecture, the outputs (pseudo IP, proto zero-phase PSF, and residual PSF), and the seismic modeling stage, which uses the generated PSF and the RC derived from the generated IP to produce modeled seismic amplitudes.

Simultaneously obtaining the PSF and IP is a non-trivial task, even for a neural network, due to the infinite combinations of different PSFs and IPs that can produce the same seismic amplitude image. To constrain the solution space of the network, we employed a regularization strategy that incorporates prior knowledge (Wu et al., 2023), guiding the outputs toward geophysically consistent solutions.

We assume that PSFs are close to zero phase. To enforce this assumption, the PSF output is divided into two components: a zero-phase PSF, which is a low-frequency component containing most of the energy, and a residual PSF, which is a complementary high-frequency component that captures finer details not represented in the zero-phase PSF.

To enforce the zero-phase assumption, we compute the autocorrelation of the proto zero-phase PSF. Autocorrelation inherently enforces the zero-phase property, aligning the PSF with geophysical constraints. The final PSF predicted by the network is reconstructed as the sum of the zero-phase PSF and the residual PSF. This approach balances the contributions of the zero-phase and high-frequency components, ensuring that the predicted PSF adheres to geophysical principles while retaining the flexibility to capture fine details.

Subsequently, we combine the LF- IP model with the generated pseudo-IP to produce the final generated IP. From the generated IP, we calculate the RC. The 2D convolutional seismic modeling process involves convolving the RC with the PSF to produce a modeled seismic amplitude. This modeled seismic amplitude is then compared with the input (true) seismic amplitude, forming a self-supervised pipeline (Figure 2).

The total loss comprises two terms: the Mean Squared Error (MSE), which measures the pixel-wise difference between the true and modeled seismic amplitudes, and the Structural Similarity Index Measure (SSIM) (Wang et al., 2004), which evaluates the similarity between the true and modeled seismic amplitudes, focusing on structural features.

The model was trained for approximately 100 epochs, with a total training time of approximately 20 minutes. A learning rate scheduler with linear decay was employed, starting with an initial learning rate of 0.003 and gradually decreasing to 0.0003 by the final epoch. We used the Adam optimizer with PyTorch's default parameters to update the network's weights. During the inference process, the reconstructed section was assembled using the TorchIO library to ensure consistency across overlapping patches. A Hanning window taper was applied to the patches to smooth transitions and avoid boundary artifacts.

**Results**

In Figure 3, we present the results of the GINN predictions after the training steps, alongside the low-frequency impedance and true amplitude for comparison. The IP (Figure 3a) and RC (Figure 3d) sections appear continuous and provide a realistic geological representation of the subsurface. Notably, as observed in conventional inversion methods, the resolution of the IP and RC sections is significantly enhanced compared to the seismic amplitude data; however, minor vertical artifacts are noticeable in the IP section. Additionally, the PSFs section (Figure 3c) produces realistic results, aligning with the expected geological illumination and demonstrating the network's capability to model spatially variable seismic resolution.

In Figures 3e and 3f, we compare the true and modeled seismic amplitudes, which directly contribute to the optimization of the network weights. The modeled seismic amplitude closely resembles the true seismic amplitude, with only minor differences, demonstrating the convergence of the GINN inversion process.

The results are highly promising. The generated amplitude sections reveal lateral variations that may not be captured by the stationary 1D average wavelet model-based inversion. However, further comparison with traditional methods for generating PSFs in model-based seismic inversion is required to comprehensively evaluate the results.

Unlike the 1D wavelet, the estimated PSFs exhibit a limited lateral resolution that effectively reduces noise and improves the accuracy of the inversion process. This feature highlights the advantages of using GINN for seismic inversion, particularly in regions with complex geological structures.

**Conclusion**

The GINN pipeline integrates deep learning techniques with seismic modeling, enabling the automatic estimation of non-stationary PSFs and the execution of model-based seismic inversion. In theory, this approach results in a more accurate approximation of acoustic impedance compared to traditional methods. The outputs, including high-resolution reflection coefficients (RC) and impedance (IP) sections, provide a detailed representation of subsurface properties.

The ability of GINN to estimate non-stationary PSFs with limited lateral resolution addresses a key limitation of the 1D stationary wavelet, which assumes an unrealistic lateral resolution. This capability not only improves the accuracy of the inversion process but also reduces noise, making the approach robust for complex geological environments.

While the results are promising, the presence of anomalies in the impedance outputs highlights areas for further investigation. Future work could focus on refining the training process, incorporating additional prior information, or exploring alternative loss functions to better constrain the solution space and minimize artifacts. Furthermore, validation with real seismic data would provide valuable insights into the applicability and performance of the proposed method in practical scenarios.

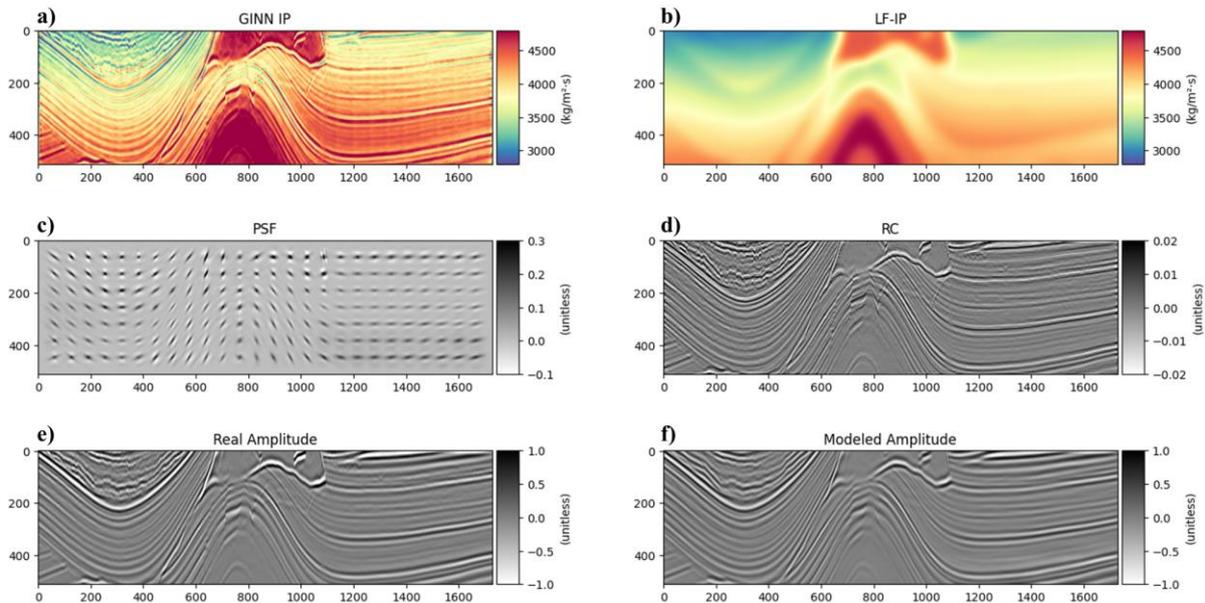

*Figure 3 Results of the GINN predictions after the training steps. (a) Impedance (IP) section, (b) low-frequency impedance (LF-IP) for comparison, (c) final Point Spread Function (PSF) section, (d) reflection coefficients (RC), (e) true seismic amplitudes, and (f) modeled seismic amplitudes.*